\documentclass[%
reprint,
superscriptaddress,
amsmath,amssymb,amsthm,mathrsfs,
aps,
prb,
]{revtex4-2}

\usepackage{graphicx}
\usepackage{dcolumn}
\usepackage{bm}
\usepackage[usenames]{color}
\usepackage{soul}
\usepackage[normalem]{ulem}
\usepackage[T1]{fontenc}
\usepackage{amsmath}

\graphicspath{{./Figures/}}
\def\TN{$T_\mathrm{N}$}

\begin{document}
	
\title{
	Probing magnetic interactions in real space via magnetic pair distribution function analysis
}
	
\author{Edison P. Carlisle}
\affiliation{ %
    Department of Physics and Astronomy, Brigham Young University, Provo, Utah 84602, USA.}

\author{Benjamin A. Frandsen}
\affiliation{ %
    Department of Physics and Astronomy, Brigham Young University, Provo, Utah 84602, USA.}
\email{benfrandsen@byu.edu}

\begin{abstract}
Determining magnetic exchange interactions is vital for understanding the magnetic behavior of materials. The standard method for exchange interaction determination involves performing inelastic neutron scattering experiments on single crystals to measure spin wave dispersions from a long-range-ordered magnetic ground state. However, the influence of exchange interactions is not limited exclusively to spin waves in ordered states, but also governs magnetic behavior in all magnetic states, including the short-range spin correlations that persist into the paramagnetic regime. Exchange parameters have been successfully recovered in the past from energy-integrated diffuse neutron scattering data from paramagnets, indicating that the diffuse scattering and underlying short-range magnetic correlations are rich in information. Consequently, magnetic pair distribution function (mPDF) analysis, which directly probes short-range magnetic correlations in real space, is expected to be well suited for extracting exchange parameters from correlated paramagnets. Here, we demonstrate a robust mean-field theory approach incorporating the Onsager reaction field for determination of exchange parameters from the results of mPDF analysis. We successfully apply this approach to several relevant materials to recover exchange interaction values comparable to literature values.
\end{abstract}
	
\maketitle

\section{Introduction}

The study of magnetic materials is central to both fundamental materials physics research and the development of novel technologies. The magnetic properties of materials are governed by magnetic exchange interactions that quantify the strength and nature of the coupling between spins. If these exchange interactions are known for a specific material, then one can often predict the material's magnetic ground state and excitations, the presence of frustration, and even the existence of exotic phases. The preferred approach for finding exchange interaction values experimentally is typically through inelastic neutron scattering measurements of spin wave dispersions. The energy- and momentum-dependent scattering data can then be modeled through spin dynamics simulations~\cite{kubo1952spin,fishman2018spin} using, for example, linear spin wave theory or the theory of $SU(N)$ coherent states ~\cite{squires1996introduction,zhang2021classical}. This approach is powerful, but it requires the material to be in an ordered state with well-defined magnon excitations. Thus, some interesting magnetic systems lacking long-range order, such as spin glasses or quantum spin liquids, cannot be directly studied with this approach.

However, inelastic neutron scattering is not the only way to probe magnetic exchange interactions. In fact, there exists a bijection between a Heisenberg magnet's thermal correlation function and its Hamiltonian~\cite{quintanilla2022relationship, murta2022one}. Thus, an alternative approach to studying long-range spin wave excitations is to study \textit{short-range} magnetic correlations in magnetic materials, which also bear the imprint of the underlying magnetic exchange interactions. Mean-field theory has emerged as a valuable method for modeling the effects of magnetic interactions in disordered states~\cite{cyrot1979new,nozieres1982onsager,reimers1991mean,logan1995onsager,wysin2000onsager,paddison2023spinteract}. Among the different variations of mean-field theory used, the Onsager reaction field (ORF) method~\cite{brout1967molecular} has proven to be straightforward and successful, often able to predict paramagnetic behavior correctly all the way down to the critical temperature~\cite{logan1995onsager,eastwood1995onsager,paddison2023spinteract}. The general principle underlying the ORF approach is that the mean-field at the site of a specific spin in the material is modified to omit the contribution of that spin itself to the field. This serves to decouple the spin from the mean-field it experiences, resulting in more accurate predictions. From the mean-field calculations, the spin correlations can be deduced and the corresponding neutron scattering pattern can be computed. This scattering is necessarily diffuse in reciprocal space, as it comes from a disordered state. Over the past few decades, exchange interactions have been studied in a number of materials using diffuse magnetic scattering rather than inelastic scattering~\cite{reimers1992diffuse,kanada2002neutron,hohlwein2002magnetic,enjalran2004theory,paddison2020scattering,paddison2023spinteract}, especially following the release of the \textsc{Spinteract} program for modeling interactions~\cite{paddison2023spinteract,alvarado2024frustrated,pitcairn2024controlling,boraley2025microscopic,desgranges2025can,potashnikov2025magnetic,paddison2026kitaev,desai2026machine}.

Although diffuse magnetic scattering can be fully calculated and described in reciprocal space, the short-range spin correlations responsible for the diffuse scattering are often more intuitively understood in real space. Consequently,  real-space treatments of diffuse magnetic scattering, e.g. magnetic pair distribution function (mPDF) analysis~\cite{frandsen2014magnetic,frandsen2015magnetic,frandsen2024new}, may also be well suited for applications of mean-field theory to extract information about magnetic exchange interactions. Experimentally, the mPDF is obtained as the Fourier transform of the magnetic structure factor and therefore provides a description of pairwise magnetic correlations as a function of real-space distance between spins. The mPDF has been instrumental in elucidating the properties in many materials, from geometrically frustrated magnets and quantum materials to functional magnetic materials and more~\cite{frandsen2024new,frandsen2020nanoscale,dun2021neutron,nuttall2023quantitative,baral2022real}. While mPDF excels at determining local magnetic structure, its potential for determining the underlying exchange interactions directly remains underexplored. Given that these interactions are bond-dependent and are generally negligible past the first few neighboring spins, and that the ORF itself is defined by local magnetic correlations (see Equation~\ref{eq:ORF}), it seems only natural that one should be able to study the exchange interactions in real space with the mPDF---an idea that we investigate here.

In this paper, we establish a mean-field theoretical framework for studying magnetic exchange interactions with mPDF. The paper proceeds as follows: Section~\ref{sec:mft} describes how real space correlations can be calculated from magnetic exchange parameters using mean-field theory. Section~\ref{sec:mpdf} reintroduces the mPDF and outlines how to extract magnetic exchange interactions from mPDF data and analysis. In Section~\ref{sec:examples}, we give several examples of materials for which we have successfully extracted exchange interactions from the results of mPDF analysis. We provide discussion and conclusions in Section~\ref{sec:conclusions}.

\section{Theory}
\subsection{Mean-Field Theory}
\label{sec:mft}
Our method of modeling the real-space magnetic correlations is based on the framework used by the other ORF techniques and involves finding the correlation function of a Heisenberg magnet. A Heisenberg magnet in the presence of a magnetic field $\mathbf{H}$ can be modeled with the following Hamiltonian:
\begin{multline}
    \mathcal{H} = -\frac{1}{2}\sum_{ij}\sum_{\alpha\beta}\sum_{\mathbf{R},\mathbf{R'}}
    J_{ij}^{\alpha\beta}(\mathbf{R}-\mathbf{R'})S_{i}^{\alpha}(\mathbf{R})S_{j}^{\beta}(\mathbf{R'})
    \\-\sum_{i,\alpha,\mathbf{R}} H_{i}^{\alpha}(\mathbf{R})S_{i}^{\alpha}(\mathbf{R}),
\end{multline}
where $\mathbf{R}$ is the position of a primitive unit cell, $i$ and $j$ label spin sites within said unit cell, $\alpha,\beta\in\{x,y,z\}$, and $\mathbf{S}$ is a spin vector. The matrix $J_{ij}^{\alpha\beta}(\mathbf{R}-\mathbf{R'})$ gives the exchange interaction between a spin at site $i$ in the unit cell at $\mathbf{R}$ and a spin at site $j$ in the unit cell at $\mathbf{R'}$. We use the convention that a positive exchange interaction prefers parallel alignment while a negative exchange interaction prefers anti-parallel alignment. We define the mean-field felt at site $\mathbf{R}$, $i$ to be
\begin{multline}
   H^{\alpha}_{i,\textrm{mf}}(\mathbf{R}) = H^{\alpha}_i(\mathbf{R}) - \lambda \langle S^{\alpha}_i(\mathbf{R}) \rangle \\ + \frac{1}{2}\sum_{j,\beta,\mathbf{R'}} J_{ij}^{\alpha\beta}(\mathbf{R}-\mathbf{R'}) \langle S_{j}^{\beta}(\mathbf{R'}) \rangle,
\end{multline}
where 
\begin{equation}
    \lambda=\sum_{j,\beta,\mathbf{R'}} J_{ij}^{\alpha\beta}(\mathbf{R}-\mathbf{R'})\frac{\langle \mathbf{S}_i^\alpha(\mathbf{R}) \mathbf{S}_j^\beta(\mathbf{R'}) \rangle}{|\mathbf{S}|^2}
    \label{eq:ORF}
\end{equation}
is the temperature-dependent Onsager Reaction Field (ORF). The inclusion of the ORF improves the accuracy of the mean-field at site $\mathbf{R}$, $i$ by removing the influence of the spin at that site from the mean-field \cite{brout1967molecular}. By treating the mean-field at $\mathbf{R}$, $i$ as independent of the spin at the same site, we can find $\langle S_i^\alpha(\mathbf{R}) \rangle$ in the standard way for an uncoupled paramagnet. For small $|\mathbf{H}_{i,\textrm{mf}}(\mathbf{R})| |\mathbf{S}|/T$, this yields
\begin{multline}
    \langle S^{\alpha}_i(\mathbf{R}) \rangle = \frac{|\mathbf{S}|^2}{d_ST}H^{\alpha}_{i,\textrm{mf}}(\mathbf{R}) 
    \\= \chi_0 \bigg ( H^{\alpha}_i(\mathbf{R}) - \lambda \langle S^{\alpha}_i(\mathbf{R}) \rangle \\
    + \frac{1}{2}\sum_{j,\beta,\mathbf{R'}} J_{ij}^{\alpha\beta}(\mathbf{R}-\mathbf{R'}) \langle S_{j}^{\beta}(\mathbf{R'}) \rangle\bigg ),
\end{multline}
where $d_S$ is the spin dimension (e.g. $d_S=1$ for Ising spins and 3 for Heisenberg spins), $\chi_0 = |\mathbf{S}|^2/d_ST$ is the Curie susceptibility, and $T$ is temperature.

We now Fourier transform to reciprocal space to simplify the above equation:
\begin{multline}
    \langle S^{\alpha}_i(\mathbf{q}) \rangle = \chi_0 \bigg ( H^{\alpha}_i(\mathbf{q}) - \lambda \langle S^{\alpha}_i(\mathbf{q}) \rangle \\ + \sum_{j,\beta} J_{ij}^{\alpha\beta}(\mathbf{q}) \langle S_{j}^{\beta}(\mathbf{q}) \rangle\bigg ),
\end{multline}
where
$S^{\alpha}_i(\mathbf{q})$, $S^{\beta}_j(\mathbf{q})$, $H^{\alpha}_i(\mathbf{q})$, and $J_{ij}^{\alpha\beta}(\mathbf{q})$ are the respective Fourier transforms of $S^{\alpha}_i(\mathbf{R})$, $S^{\beta}_j(\mathbf{R})$, $H^{\alpha}_i(\mathbf{R})$, and $J_{ij}^{\alpha\beta}(\mathbf{R}-\mathbf{R'})$. Additionally, we diagonalize $J_{ij}^{\alpha\beta}(\mathbf{q})$, with eigenvalues and eigenvectors $J_\mu(\mathbf{q})$ and $U^{\alpha}_{i\mu}(\mathbf{q})$ defined by
\begin{equation}
    J_\mu(\mathbf{q})U^{\alpha}_{i\mu}(\mathbf{q}) = \sum_{j\beta}J_{ij}^{\alpha\beta}(\mathbf{q})U^{\beta}_{j\mu}(\mathbf{q}),
\end{equation}
and perform the rest of the analysis in the eigenbasis. Now we solve for the average value of the spin:
\begin{equation}
    \langle S_\mu (\mathbf{q}) \rangle = \frac{\chi_0}{1-\chi_0(J_\mu(\mathbf{q})-\lambda)}H_\mu (\mathbf{q})
\end{equation}
or 
\begin{equation}
    \chi_\mu (\mathbf{q}) = \frac{\chi_0}{1-\chi_0(J_\mu(\mathbf{q})-\lambda)}.
\end{equation}
Not only do the eigenvalues of the exchange matrix ameliorate the math, they also contain useful physics. For example, the maximum eigenvalue is directly related to the ground state energy through $E_\mathrm{ground} = - \mathrm{max}_{\mu,\mathbf{q}}\{J_\mu(\mathbf{q})\}|\mathbf{S}|^2$. For ordered systems the corresponding $\mathbf{q}$ is the magnetic ordering vector. However, the presence of frustration complicates the magnetic ordering vector. Fully frustrated systems may have a flat $J_\mu(\mathbf{q})$, signaling a total lack of ordering~\cite{reimers1991mean,canals2000mean}.

According to the fluctuation-dissipation theorem, at high temperatures $\chi_\mu(\mathbf{q}) = \frac{1}{T} \langle S_\mu(\mathbf{q})S_\mu(\mathbf{-q}) \rangle$. Consequently, the reciprocal-space correlation function can be expressed as
\begin{equation}
    \langle S_\mu(\mathbf{q})S_\mu(\mathbf{-q}) \rangle = T\frac{\chi_0}{1-\chi_0(J_\mu(\mathbf{q})-\lambda)}.
    \label{eq:SpinCorrQ}
\end{equation}
The previous derivations have been used successfully to extract exchange interactions from reciprocal-space diffuse scattering data~\cite{logan1995onsager,eastwood1995onsager,hohlwein2002magnetic,paddison2023spinteract}. Here we divert from previous studies by focusing primarily on the real-space correlations. 

For an exchange interaction $J_\mu(\mathbf{q})$ that is isotropic with respect to $\mathbf{q}$, we can get a simple look at the real-space spin correlations if we approximate $\langle S_\mu(\mathbf{q})S_\mu(\mathbf{-q}) \rangle$ as a series of Lorentzian peaks
\begin{equation}
    \langle S_\mu(\mathbf{q})S_\mu(\mathbf{-q}) \rangle \approx T\sum_{\mathbf{q}_0} \frac{A}{\xi_\mu^{-2}+|\mathbf{q}-\mathbf{q}_0|^2},
\end{equation}
where $A$ is some constant and $\mathbf{q}_{0}$ corresponds to a maximum of $J_\mu(\mathbf{q})$ (note that this is only for the index $\mu$ that also corresponds to a maximum of $J_\mu(\mathbf{q})$). This is easily done by assuming small $\mathbf{q}-\mathbf{q}_0$ so that 
\begin{equation}
    J_\mu(\mathbf{q}) \approx J_\mu(\mathbf{q}_0) + \frac{1}{2} J_{\mu}^{\prime\prime}\delta q^2,
\end{equation}
where we define $\delta q = |\mathbf{q}-\mathbf{q}_0|$ and $J_{\mu}^{\prime\prime} =  \frac{\partial^2J_\mu}{\partial q^2}(\mathbf{q}_0)$. Then,
\begin{equation}
    \langle S_\mu(\mathbf{q})S_\mu(\mathbf{-q}) \rangle \approx T
    \sum_{\mathbf{q}_0} \frac{-2/J_{\mu}^{\prime\prime}}{\xi_\mu^{-2}+\delta q^2},
    \label{eq:SpinCorrLorentzian}
\end{equation}
\begin{equation}
    \xi = \sqrt{\frac{-\frac{1}{2}J_{\mu}^{\prime\prime}}{1/\chi_0-J_\mu(\mathbf{q}_0) + \lambda}}.
\end{equation}

We transform back to real space through means of an integral in reciprocal space rather than a sum over wavevectors. As we are now dealing with only a Lorentzian, we integrate over all of reciprocal space rather than just within the first Brillouin Zone (BZ) under the assumption that the contributions of the Lorentzian outside the BZ are too small to significantly affect the results. The real space correlation function is therefore given by
\begin{equation}
    \langle S_\mu(0) S_\mu(\mathbf{r}) \rangle \approx \frac{V_\mathrm{UC}T}{2\pi}\sum_{\mathbf{q}_0} \frac{\exp{(-|\mathbf{r}|/\xi)}}{J_{\mu}^{\prime\prime}|\mathbf{r}|}e^{-i\mathbf{q}_0\cdot\mathbf{r}},
    \label{eq:SpinCorrIso}
\end{equation}
where $V_{\mathrm{UC}}$ is the volume of the unit cell and $\mathbf{r}$ is the conjugate variable of $\mathbf{q}$. $\xi$ thus becomes the correlation length and $\mathbf{q}_0$ is the ordering vector, as expected.

Ref.~\cite{frandsen2022diffpy} defines anisotropic correlation lengths through an ellipsoid. This ellipsoid is defined by $\mathbf{x}^T D \mathbf{x} = 1$ where $x$ is the distance from the center of the ellipsoid to the surface and $D$ is a matrix with eigenvalues corresponding to the inverse squares of the semi-axes lengths and eigenvectors corresponding to semi-axes directions. In the context of anisotropic correlation lengths, we call $D$ the damping matrix. The correlation length in direction $\hat{n}$ can be represented as $\xi = (\hat{n}^TD\hat{n})^{-1/2}$. By performing the Fourier transform of $\langle S_\mu(0) S_\mu(\mathbf{r}) \rangle$ for an anisotropic $J_\mu(\mathbf{q})$, still assuming small $\delta q$, we get 
\begin{equation}
    \langle S_\mu(0)S_\mu(\mathbf{r}) \rangle \approx \frac{V_\mathrm{UC}T}{2\pi}\sum_{\mathbf{q}_0}\frac{e^{-i\mathbf{q}_0\cdot\mathbf{r}}}{\sqrt{-\det C}} 
    \frac{\exp(- \sqrt{\mathbf{r}^T D~\mathbf{r}})}
    {\sqrt{-\mathbf{r}^T C^{-1}~\mathbf{r}}},
    \label{eq:SpinCorrMuAniso}
\end{equation}
where $C_{\alpha\beta} = \partial_{q_\alpha}\partial_{q_\beta} J_\mu(\mathbf{q}_0)$. Through the Fourier transform, we find the damping matrix $D$ through the expression
\begin{equation}
    [D^{-1}]_{\alpha\beta} = \frac{-\frac{1}{2}\partial_{q_\alpha}\partial_{q_\beta} J_{\mu}(\mathbf{q}_0)}{1/\chi_0-J_\mu(\mathbf{q}_0) + \lambda}.
    \label{eq:DampingMatrix}
\end{equation}
These results resolve to Eq.~\ref{eq:SpinCorrIso} for isotropic $J_\mu(\mathbf{q})$. The derivation of these results, as well as results for the case of a 1D or 2D magnet, can be found in Appendix~\ref{sec:appendixFT}.

While a rigorous derivation of $\langle S^\alpha_i(0)S^\beta_j(\mathbf{R}) \rangle$ would require a convolution of $\langle S_\mu(0)S_\mu(\mathbf{r}) \rangle$ with the Fourier transformed eigenvectors $U^{\alpha}_{i\mu}(\mathbf{q})$, we take a more approximate shortcut. In the limit of $T \rightarrow 0$, Equation~\ref{eq:SpinCorrLorentzian} approaches a series of Dirac delta functions. This represents the ordered state of the system rather than the paramagnetic state. Then, when calculating $\langle S^\alpha_i(-\mathbf{q})S^\beta_j(\mathbf{q})\rangle$, we no longer need to know the eigenvectors at each $\mathbf{q}$ as they will only be relevant at each $\mathbf{q}_0$ corresponding to a delta function peak. Then, in real space, we need only use constant eigenvectors $U^\alpha_{i\mu}(\mathbf{q}_0)e^{i\mathbf{q_0}\cdot\mathbf{r}_i}$ where $\mathbf{r}_i$ is the position of the $i$th magnetic atom in the unit cell. Thus, we approximate $\langle S^\alpha_i(0)S^\beta_j(\mathbf{R}) \rangle$ as
\begin{multline}
    \langle S^\alpha_i(0)S^\beta_j(\mathbf{R}) \rangle \approx \frac{V_\mathrm{UC}T}{2\pi N}\sum_{\mu,\mathbf{q}_0}\frac{e^{-i\mathbf{q}_0\cdot\mathbf{R}}}{\sqrt{-\det C}} 
    \\ \times \frac{\exp(- \sqrt{\mathbf{r}^T D~\mathbf{r}})}
    {\sqrt{-\mathbf{r}^T C^{-1}~\mathbf{r}}}\frac{U^{\alpha}_{i\mu}(\mathbf{-q_0})U^{\beta}_{j\mu}(\mathbf{q_0})}{|U^{\alpha}_{i\mu}(\mathbf{-q_0})||U^{\beta}_{j\mu}(\mathbf{q_0})|},
    \label{eq:SpinCorrAniso}
\end{multline}
where now $\mathbf{r} = \mathbf{R} + \mathbf{r}_j - \mathbf{r}_i$, $N$ is equal to the number of magnetic atoms in the primitive unit cell, and, as before, we only consider the indices $\mu$ for which a maximum of $J_\mu(\mathbf{q})$ occurs. This approximation can often communicate the essential spin correlations of a material without the need for calculating the eigenvalues everywhere in the first BZ and performing the subsequent convolutions. However, we stress here that while we rely on a low-temperature limit for this approximation, this method is only reliable in the paramagnetic regime, particularly at high temperature, though we still find good agreement in the critical regime (see Section~\ref{sec:examples}).

We have yet to address how to calculate the ORF. While a popular method of calculation the ORF is constraining it through a sum rule in reciprocal space of the form~\cite{logan1995onsager,eastwood1995onsager,hohlwein2002magnetic,paddison2023spinteract}
\begin{equation}
    \sum_{\mu,\mathbf{q}} \langle S_\mu(\mathbf{q})S_\mu(-\mathbf{q}) \rangle = N_\mathbf{q}N|\mathbf{S}|^2,
\end{equation}
where $N_\mathbf{q}$ is the number of $\mathbf{q}$ points in the first BZ, we elect to calculate the ORF through the original definition in Equation~\ref{eq:ORF}, where $\langle \mathbf{S}_i^\alpha(\mathbf{R}) \mathbf{S}_j^\beta(\mathbf{R'}) \rangle$ is now dependent on the ORF. This leads to a transcendental equation that allows us to solve for the ORF quickly and self-consistently at each temperature.

\subsection{mPDF}
\label{sec:mpdf}
Having a real-space description of exchange-dependent spin correlations is only useful when real-space spin correlations can be measured. Over the past decade, the magnetic pair distribution function (mPDF) has proven to be a robust way to probe these correlations~\cite{frandsen2024new}. The mPDF is the Fourier transform of the total magnetic scattering, and therefore includes both the long-range correlation information contained in magnetic Bragg peaks as well as the short-range correlation information contained in the diffuse magnetic scattering. This makes mPDF an excellent probe of the paramagnetic regime, in which short-range correlations dominate the magnetic scattering. For an isotropic powder sample with a single spin-only magnetic species, the mPDF is given by~\cite{frandsen2014magnetic,kodama2017alternative}
\begin{widetext}
    \begin{align}
        \textrm{mPDF}(r) &= \frac{2}{\pi}\int_{q_\textrm{min}}^{q_{\textrm{max}}} q \left ( \frac{(\textrm{d}\sigma/\textrm{d}\Omega)_\textrm{mag}}{\frac{2}{3}N_sS(S+1)(\gamma r_0)^2[f(q)]^2}-1 \right )\sin (qr) \textrm{d}q \\
        &= \frac{3}{2S(S+1)}\left ( \frac{1}{N_s}\sum_{i \neq j}\left [ \frac{A_{ij}}{r}\delta(r-r_{ij}) + B_{ij} \frac{r}{r_{ij}^3}\Theta(r_{ij}-r) \right ] - 4\pi\rho_0\frac{2}{3}m^2 \right ).
        \label{eq:mPDFTheory}
    \end{align}
\end{widetext}
The first equation shows how to generate the mPDF from experimental data and the second equation defines the mPDF theoretically for a given magnetic structure. Here, $r$ is the real-space distance, $q$ is the scattering vector magnitude, $q_\textrm{min}$ and $q_\textrm{max}$ are the smallest and largest measured $q$ values respectively, $(\textrm{d}\sigma/\textrm{d}\Omega)_\textrm{mag}$ is the magnetic differential scattering cross section, $N_s$ is the number of magnetic spins in the system, $S$ is the spin quantum number in units of $\hbar$, $\gamma=1.913$ is the neutron magnetic moment in units of nuclear magnetons, $r_0 = (\mu_0e^2)/(4\pi m_e)$ is the classical electron radius, $f(q)$ is the magnetic form factor, $i$ and $j$ now index all spins within the system rather than just spins within the unit cell, $r_{ij}$ is the distance between two spins, $A_{ij} = \langle S_i^yS_j^y \rangle$, $B_{ij} = 2\langle S_i^xS_j^x\rangle - \langle S_i^yS_j^y \rangle$, $\Theta$ is the Heaviside step function, $\rho_0$ is the number of spins per unit volume, and $m$ is the average magnetic moment in Bohr magnetons. For $A_{ij}$ and $B_{ij}$, a local coordinate system is used, with $\hat{x} = (\mathbf{r}_j-\mathbf{r}_i)/|\mathbf{r}_j-\mathbf{r}_i|$ and $\hat{y} = [\mathbf{S}_i - \hat{x}(\mathbf{S}_i\cdot\hat{x})]/|\mathbf{S}_i - \hat{x}(\mathbf{S}_i\cdot\hat{x})|$.

For a material with multiple magnetic species or with orbital contributions to the magnetic moment, we substitute $S(S+1)$ in Equation~\ref{eq:mPDFTheory} with $\left\langle g[J(J+1)]^{1/2} \right\rangle^2$ and substitute $A_{ij}$, $B_{ij}$ with $g_i g_jA_{ij}$, $g_i g_jB_{ij}$, where $g$ is the Land\'e splitting factor, $J$ is now the total angular momentum (though elsewhere in the paper we assume $S$ could be the spin or total angular momentum and reserve $J$ for the magnetic exchange interaction), and $\langle ... \rangle$ denotes an average over all magnetic moments in the system~\cite{frandsen2014magnetic}.

In practice, fitting a theoretical mPDF to paramagnetic experimental data through the Python package \textit{diffpy.mpdf}~\cite{frandsen2022diffpy} includes determination of the magnetic correlation length or damping matrix, orientations of spins within the unit cell, and the magnetic ordering vector. These are all quantities that can be determined from the ORF approach to calculating spin correlations. Additionally, through mPDF analysis one can calculate the short-range ordered moment, also called the local magnetic order parameter, defined by $m_\mathrm{SRO}=g\sqrt{|\langle \mathbf{S}(0)\cdot \mathbf{S}(\mathbf{r}_{\mathrm{NN}})\rangle|}$, where $\mathbf{r}_\mathrm{NN}$ is the nearest neighbor distance. This quantity can easily be calculated through Equation~\ref{eq:SpinCorrAniso}. Thus, we suggest that using paramagnetic mPDF data in tandem with ORF-based mean-field calculations can be a useful approach to determining magnetic exchange interactions. We take the following approach in the subsequent examples: 

\begin{enumerate}
    \item Determine the material's atomic structure and the basic characteristics of the magnetic moments (lattice parameters, atom types and positions, spin quantum numbers of magnetic atoms, etc.). When determining $|\mathbf{S}|^2$, we recommend using $S^2$ instead of $S(S+1)$, because a given spin correlation $\langle S_i S_j \rangle$ cannot exceed $S^2$ in the absence of entanglement.
    \item Perform standard mPDF fits to the data in the paramagnetic regime, and extract the damping matrix and $m_\mathrm{SRO}$ from the fit results. It is helpful to have results for at least two temperatures in the paramagnetic phase, one in the critical regime near the phase transition and the other at significantly higher temperature (see Sec.~\ref{sec:conclusions} for further discussion of temperature selection). 
    \item Make an initial guess for exchange interaction parameters. Fourier transform and diagonalize the exchange interactions. Determine peak positions in $J_\mu(\mathbf{q})$.
    \item Using Eqs.~\ref{eq:DampingMatrix} and \ref{eq:SpinCorrAniso}, perform least-squares minimization or some other optimization routine to fit the exchange parameters to the experimentally determined damping matrix and $m_\mathrm{SRO}$ parameters from Step 2. It may be necessary to reevaluate the locations of the peaks of $J_\mu(\mathbf{q})$ and refit as the parameter space of the exchange interactions is explored.
\end{enumerate}

\section{Example Materials}
\label{sec:examples}
To showcase the utility of this approach, we present an analysis of three materials, each with unique properties. The data and python scripts used for these examples are available at Ref.~\cite{carlisle2026data}.

\begin{table}
    \centering
    \caption{Exchange interactions in Kelvin.}
    \begin{tabular}{lcccc}
        & J1 & J2 & J3 & J4 \\
        \hline
        \multicolumn{5}{l}{\rule{0pt}{4ex}\uline{Zn$_{0.5}$Mn$_{0.5}$Te}} \\
        Ref.~\cite{furdyna1987magnetic} & -9.5 & -1.9 & - & - \\
        \textsc{Spinteract} & -8.87$\pm$0.03 & -1.08$\pm$0.02 \\
        Fit & -10.5$\pm$1.0 & -1.0$\pm$0.1 & - & - \\
        \multicolumn{5}{l}{\rule{0pt}{4ex}\uline{MnTe}} \\
        Ref.~\cite{szusz_prb06} & -43.0$\pm$0.6 & 1.34$\pm$0.1 & -5.74$\pm$0.08 & - \\
        Ref.~\cite{mu2019phonons} & -35.6 & -0.32 & -4.64 & -1.86 \\
        Fit & -35.4$\pm$2.3 & -1.9$\pm$2.5 & -6.8$\pm$1.4 & - \\
        \multicolumn{5}{l}{\rule{0pt}{4ex}\uline{NaMnO$_2$}} \\
        Ref.~\cite{zorko2008magnetic} & -65 & -28.6 & - & - \\
        Ref.~\cite{dally2018amplitude} & -71.5 & -8.9 & - & - \\
        Fit & -65.4$\pm$1.8 & -23.9$\pm$0.5 & - & -
    \end{tabular}
    \label{tab:Js}
\end{table}

\begin{figure*}
    \centering
    \includegraphics[width=\textwidth]{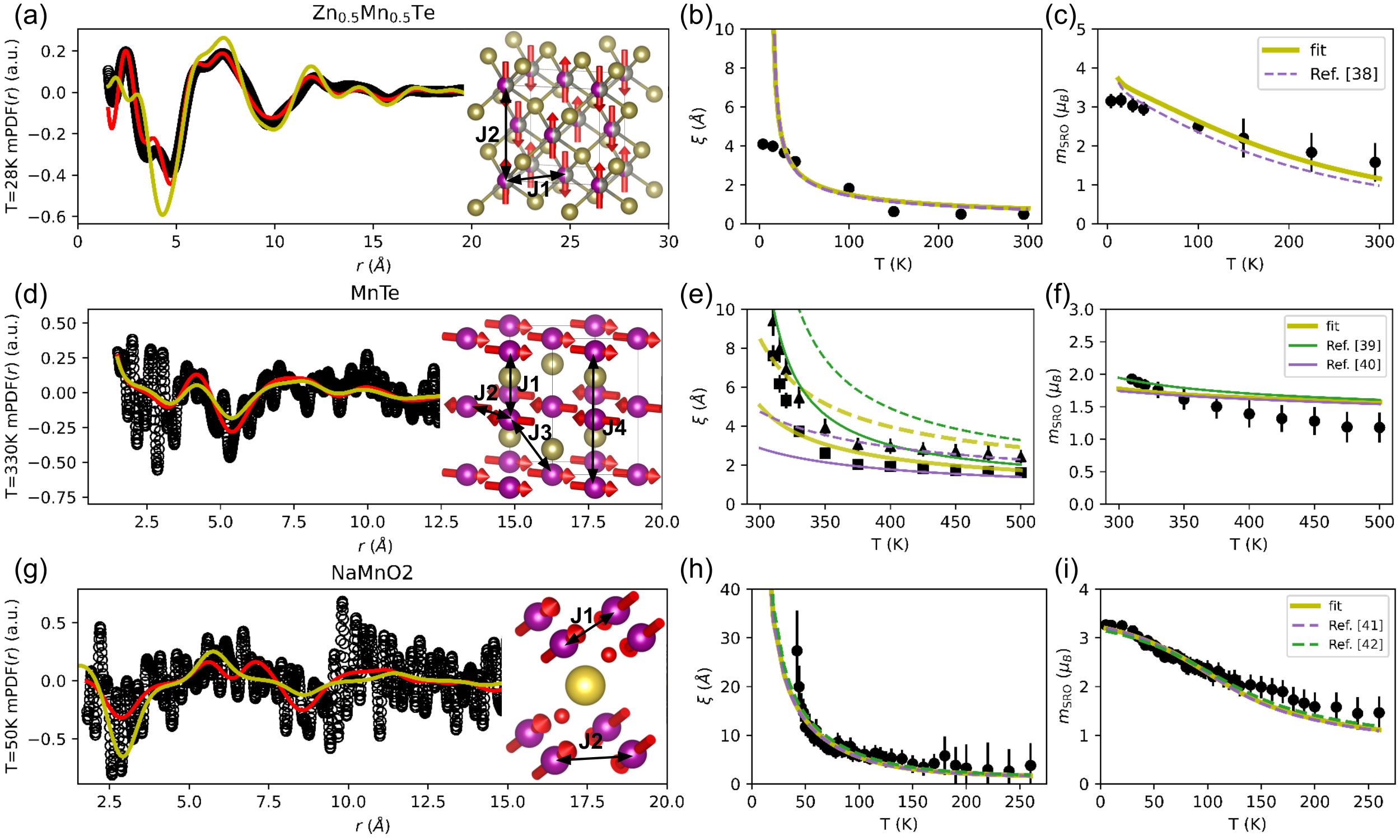}
    \caption{(a-c) Fit of Zn$_{0.5}$Mn$_{0.5}$Te mPDF data. (a) The mPDF data at 28 K is in black, with the atomistic fit~\cite{hatt2025cluster} in red and our reconstruction of the mPDF from our exchange interaction fits in yellow. (inset) The crystal structure and exchange interactions are shown. The image of the crystal structure was generated with VESTA~\cite{momma2011vesta}. (b) The correlation lengths extracted from the mPDF fits as a function of temperature are shown in black. Our fit to the data, as well as calculated correlation lengths from exchange interactions provided by Ref.~\cite{furdyna1987magnetic}, are plotted as lines. (c) The $m_\mathrm{SRO}$ extracted from the mPDF is plotted as black circles, with our fit and calculated values from the same reference are plotted as lines. (d-f) Same as (a-c), but for MnTe. (d) The 330 K mPDF data and fit are from Ref.~\cite{baral2022real}. (e,f) Dotted lines and triangles represent out of plane correlation lengths, while solid lines and squares signify in-plane correlation lengths. The green and purple lines are calculated from interaction strengths taken from Ref.~\cite{szusz_prb06} and Ref.~\cite{mu2019phonons} respectively. (g-i) Same as (a-c) but for NaMnO$_2$. (g) The 50 K mPDF data and fit are from Ref.~\cite{frandsen2020nanoscale}. (h,i) Purple and green lines are calculated from interaction strengths from Ref.~\cite{zorko2008magnetic} and Ref.~\cite{dally2018amplitude} respectively.}
    \label{fig:fits}
\end{figure*}

\subsection{Zn$_{0.5}$Mn$_{0.5}$Te}

Spin glass systems have no long-range-ordered magnetic state. This complicates the study of their magnetic interactions through the usual inelastic neutron scattering approach, as there are no conventional magnon modes to observe. Thus, to study the interactions with neutron scattering, a diffuse scattering analysis may prove useful. Recent work has solved the short-range magnetic structure of the cluster spin glass Zn$_{0.5}$Mn$_{0.5}$Te through mPDF and reverse Monte Carlo analyses of the diffuse scattering~\cite{hatt2025cluster}. From the mPDF analysis, values for the magnetic correlation length and $m_\mathrm{SRO}$ have been determined for a series of temperatures both above and below the freezing temperature $T_f$, which is between 21 and 24 K. This information is sufficient to extract the exchange interactions.

The exchange interactions of Zn$_{0.5}$Mn$_{0.5}$Te have previously been deduced from a combination of magnetic susceptibility measurements (to determine $J_1$) and theory (to determine $J_2$ from $J_1$)~\cite{furdyna1987magnetic}, giving a good baseline with which we compare our results. This cluster spin glass crystallizes into a zinc blende structure with Zn and Mn randomly distributed in a face centered cubic lattice (see Fig.~\ref{fig:fits} a) inset). Only the Mn ions are magnetic, with $S_\mathrm{Mn2+}=5/2$ and $|\mathbf{S}|^2_\mathrm{Mn2+} = S(S+1)=35/4$~\footnote{While we would suggest using $|\mathbf{S}|^2_\mathrm{Mn2+} = S^2 = 25/4$, Ref.~\cite{furdyna1987magnetic} uses $S(S+1)$ in their calculation, so we elect to maintain consistency as we compare values of exchange parameters}. To account for only half-occupancy on the magnetic sites, we use $|\mathbf{S}|^2 = \frac{1}{2}|\mathbf{S}|^2_{\mathrm{Mn2+}} = 35/8$ in our calculations of the exchange interactions. We proceeded to calculate the exchange interactions following the approach outlined in Section~\ref{sec:mpdf}. We also calculated the expected correlation lengths and short-ranged ordered moments from the literature values of the exchange interactions. Additionally, because the original mPDF study of this material in Ref.~\cite{hatt2025cluster} involved isolating the magnetic diffuse scattering, we performed a \textsc{Spinteract}~\cite{paddison2023spinteract} fit as well. In Table~\ref{tab:Js}, we see close agreement between the literature values, \textsc{Spinteract} fit values, and our fit values of the exchange interactions. We also observe agreement between the correlation lengths and $m_\textrm{SRO}$ values calculated from literature exchange interactions and mPDF data in Fig.~\ref{fig:fits} (b-c). The calculated correlation length (Fig.~\ref{fig:fits} (b)) diverges at around 20 K, corresponding reasonably well to the observed values of $T_f$, which fall in the range 21 -- 24~K.

\subsection{MnTe}

Hexagonal MnTe has recently garnered attention as a leading example of an altermagnet~\cite{gonzalez2023spontaneous,mazin2023altermagnetism,lee2024broken,krempasky2024altermagnetic}, a newly discovered phase of spin-compensated collinear magnets with spin-split electronic bands without the need for spin-orbit coupling~\cite{vsmejkal2022beyond}. Additionally, MnTe has found applications in spintronics~\cite{kriegner2016multiple,bossini2021femtosecond}, when doped can be an effective thermoelectric~\cite{xu2017performance,ren2017synergistic,dong2018lead,deng2021high,xiong2023lattice,zulkifal2023multiple} due to strong short-range magnetic correlations above the N\'eel temperature (\TN)~\cite{zheng2019paramagnon,polash2021understanding,baral2022real}, and experiences strong magnetostructural coupling due to the bond-length dependence of the exchange interaction strength~\cite{baral2023,carlisle2025tuning}. As shown in the inset of Fig.~\ref{fig:fits} d), the magnetic structure of MnTe is such that its spins are aligned within the $ab$-plane and anti-aligned between planes. Above \TN, the correlation length is anisotropic, with stronger correlations out of plane than in plane~\cite{baral2022real}. These effects are due to a strong antiferromagnetic nearest-neighbor exchange interaction along the $c$-axis~\cite{szusz_prb06,mu2019phonons}. The anisotropic nature of the correlations make MnTe an excellent case study for extracting exchange interactions from real-space data.

The crystal structure of MnTe belongs to space group $P6_3/mmc$~\cite{johnston1961mnte}. The Mn$^{2+}$ ions have $S=5/2$ so $|\mathbf{S}|^2 = 25/4$. From this preliminary information, we can begin to convert from real-space data to exchange interactions and vice versa. We first calculated the expected correlation lengths and $m_\mathrm{SRO}$ for exchange interactions taken from Refs.~\cite{szusz_prb06} and~\cite{mu2019phonons}, which we compare in Fig.~\ref{fig:fits}(e,f) to the values determined experimentally from Ref.~\cite{baral2022real}. The exchange values from Ref.~\cite{szusz_prb06} correctly predict that the correlation lengths increase dramatically near the true \TN~ of 307 K~\cite{walther1967}, but the agreement at high temperature is somewhat poor. In contrast, the exchange values from Ref.~\cite{mu2019phonons} provide good agreement at high temperature but fail to capture the behavior in the critical regime. We then performed our own fit of the exchange interactions to the correlation length and $m_\mathrm{SRO}$ data. We included interactions up to $J_3$ but excluded $J_4$, because $J_4$ introduced degeneracies between the exchange parameters in our fitting algorithm, likely because $J_4$ is along the $c$-axis and therefore affects the system in a similar way to $J_1$. The fit correctly identified the order vector as (0,0,0), and the eigenvectors tell us that the two spins within the unit cell are anti-aligned, as observed experimentally. For the correlations calculated from the exchange values from Ref.~\cite{szusz_prb06}, we also included the easy-plane anisotropy. This did not affect the calculated values of the correlation lengths or $m_\textrm{SRO}$, but correctly predicted that the preferred orientation of the spins was in-plane. Our fitted exchange interactions are compared to those from Refs.~\cite{szusz_prb06} and~\cite{mu2019phonons} in Table~\ref{tab:Js}, showing reasonable consistency. From Fig.~\ref{fig:fits}(e,f), we see that our fit lies between the two sets of curves produced by the literature results, providing better overall agreement between the critical regime and the high-temperature asymptote than the other sets of interaction values.

\subsection{NaMnO$_2$}
In geometrically frustrated magnets, the interplay between the position of the spins on the lattice and the nature of the magnetic  interactions (e.g. ferromagnetic or antiferromagnetic) is such that there is no spin configuration that fully satisfies the magnetic interactions. This can lead to exotic ground states such as quantum spin liquids. Although the frustration can suppress the magnet's ability to develop long-range order, the material will still often manifest short-range magnetic order~\cite{frandsen2024new,ramirez2025short}. The frustrated antiferromagnetic NaMnO$_2$ has magnetic atoms situated in an isosceles triangle configuration with one short edge and two slightly longer edges and enters an ordered state below \TN~$ = 45$ K ~\cite{giot2007magnetoelastic}. Above \TN, the short-range magnetic order is quasi-one-dimensional, with robust antiferromagnetic short-range correlations along a chain in the nearest-neighbor direction and much more suppressed short-range order in other directions~\cite{stock2009one,dally2018amplitude,dally2018thermal,frandsen2020nanoscale}. This quasi-one-dimensional behavior arises due to the frustration of the $J_2$ interaction. No configuration of the second-nearest-neighbor spins satisfies the exchange interaction, so correlations do not persist in the $J_2$ direction. This gives NaMnO$_2$ more one-dimensional behavior in the paramagnetic phase than even MnTe, which has a $J_1$ interaction that is significantly stronger than all of its other interactions but still has correlations that persist in directions orthogonal to the $J_1$ interaction. Therefore, NaMnO$_2$ is a well suited candidate for simultaneously testing our method on both frustrated magnets and quasi-one-dimensional magnets.

The mPDF data we used for this material were published in Ref.~\cite{frandsen2020nanoscale}. We used the 1D version of the spin correlation function found in Equation~\ref{eq:SpinCorr1D}. We used two separate sources for exchange interaction strengths to compare to our methods: one from magnetization measurements~\cite{zorko2008magnetic} and one from inelastic neutron scattering~\cite{dally2018amplitude}. The magnetic ions in NaMnO$_2$ are Mn$^{3+}$ ($S=4/2$), so we would expect $|\mathbf{S}|^2 = 4$. However, when calculating spin correlations from literature exchange interactions, both of these cases produced $m_\mathrm{SRO}$ values far exceeding the results from the mPDF results. In fact, the ordered moment at 4 K is only $2.92~\mu_B$, smaller than the expected $4~\mu_B$ for $S=4/2$ spins~\cite{giot2007magnetoelastic}. We suggest that there may be a reduction of the bare ordered moment of the Mn atoms. To account for this, we included $|\mathbf{S}^2|$ as an empirical fitting parameter. We fit our model to the correlation length data from \TN~$=45$ K upward and to the $m_\mathrm{SRO}$ data at all available temperatures. From our fitting, we found exchange values as listed in Table~\ref{tab:Js} and $|\mathbf{S}|^2 = 2.4\pm0.1$, which is equivalent to a bare ordered moment of $3.1~\mu_B$. This value is not far off from the observed $m_\mathrm{SRO}$ in Fig.~\ref{fig:fits} (i). For our calculations of the correlation lengths and $m_\mathrm{SRO}$ based on literature exchange values, we fit $|\mathbf{S}|^2$ to account for the reduction of the moment and got $|\mathbf{S}|^2 = 2.4$ for Ref.~\cite{zorko2008magnetic} and 2.5 for Ref.~\cite{dally2018amplitude}. We see in Fig.~\ref{fig:fits} (g-i) that the fits, data, and calculations from literature exchange values all agree very well. In fact, Fig.~\ref{fig:fits} (i) shows agreement even well below \TN~in $m_\mathrm{SRO}$. Interestingly, $J_2$ varies dramatically between the two literature values with only small differences in correlation length and $m_\mathrm{SRO}$ values. However, this model is still sensitive enough to $J_2$ that the goodness of fit was significantly better for a $J_2$ near $-25$ K than a $J_2$ near $-5$ K. Changing $J_2$ also has a small affect on the position of the maximum $J_\mu(\mathbf{q})$. For $J_2=0$, $\mathbf{q}_0 = (1/4, -1/4, 0)$, which is along the nearest neighbor direction and corresponds to an antiferromagnetic chain. As the magnitude of $J_2$ increases, $\mathbf{q}_0$ moves slightly off $(1/4, -1/4, 0)$. This shift signifies the presence of frustration, but the overall magnetic structure remains qualitatively similar.

\section{Discussion \& Conclusions}
\label{sec:conclusions}

We have demonstrated the successful extraction of magnetic exchange interactions using mPDF analysis, bringing an important new capability to the growing suite of mPDF methods. While inelastic neutron scattering remains a crucial reference point for determining magnetic exchange interactions, the mPDF-based approach introduced here can provide a valuable alternative. Magnetic PDF data can be obtained simultaneously with standard atomic PDF data on modern neutron PDF instruments, which can produce high-quality data even from small powder samples that may not be suitable for inelastic neutron scattering measurements. The three-fold combination of atomic PDF analysis to probe the local atomic structure, standard mPDF fits to determine the short-range magnetic structure, and this new analysis method to determine exchange parameters—--all performed on the same data set---offers a powerful and unique view into magnetic materials, particularly complex magnets with important interplay between structure and magnetism. This mPDF-based approach for inferring exchange parameters can also be used as a stand-alone method separate from atomic PDF analysis, e.g. on data collected from neutron diffractometers not specifically optimized for atomic PDF~\cite{calder2025investigating} or from polarized neutron diffractometers~\cite{frandsen2022magnetic}. In this case, performing the fits and analysis in real space after Fourier transformation is a useful complement to reciprocal-space-based approaches like \textsc{Spinteract}~\cite{nuttall2023quantitative} or a recent machine-learning algorithm~\cite{desai2026machine}. It should also be noted that this method is not limited to use in mPDF analysis. If the magnetic correlation length or $m_\textrm{SRO}$ can be experimentally deduced by some other method, such as from diffuse scattering in reciprocal space or magnetometry measurements, this technique could still be used.

When planning mPDF experiments to probe exchange interactions, identifying the optimal temperatures for data collection is essential. To determine which temperatures are most important for model fitting, we applied an optimal experiment design algorithm~\cite{kurniawan2026information} to each of our three materials to find the minimal temperature set necessary to constrain the exchange parameters. For each material, only two temperatures were required by the algorithm: one in the critical regime where the correlation length is long, and another further into the paramagnetic regime after the correlation length has started to asymptote. Additional measurements at higher temperatures can confirm that the correlation length has asymptoted. Because this method is a high-temperature approximation that is not guaranteed to accurately model the critical regime, we also recommend collecting data at a temperature between the asymptotic region and the low-temperature critical regime.

We now discuss limitations of this method. By using powder neutron diffraction data, the directional dependence of some exchange interactions may be difficult to resolve due to the spherical averaging characteristic of powder data. However, the same principles presented here could be applied to 3D-$\Delta$mPDF~\cite{roth2018model} data to better probe directional dependence. Another point to observe is that this method relies on approximating the $\mathbf{q}$-dependent magnetic susceptibility as Lorentzian-like peaks. Thus, we are potentially losing information that would help constrain the exchange interaction values for more complex materials. However, this brings the benefit of simpler model-fitting, as we only have to fit to the the damping matrix and $m_\mathrm{SRO}$ as opposed to fitting to the magnetic scattering at each measured reciprocal space point for each temperature.

This work invites further research in several directions. This method would benefit from further testing on a variety of different materials (e.g. materials with strong  Dzyaloshinskii-Moriya interactions) to probe other strengths and weaknesses. Another interesting point of research would be modeling the entire mPDF through mean-field theory instead of just identifying key parameters such as the damping matrix, $m_\mathrm{SRO}$, ordering vector, and relative spin orientation. Finally, adapting this method specifically to 3D-$\Delta$mPDF data is likely to improve accuracy by taking into account the full directionality of the magnetic correlations. Even without these future developments, this work highlights the value of mPDF methods not only for revealing details of short-range magnetic order, but also yielding insights into the fundamental interactions between the spins themselves.

\textbf{Data Availability}
The data that support the findings of this article are openly available~\cite{carlisle2026data}.

\textbf{Acknowledgements}
We thank Joe Paddison for helpful discussions about extracting exchange interactions from magnetic diffuse scattering data. The theory and modeling work in this paper were supported by the United States National Science Foundation, Division of Materials Research, LEAPS-MPS program through Award No. 2418438.

\textbf{Competing Interests}
All authors declare no financial or non-financial competing interests.

\bibliography{interactions}

@article{alvarado2024frustrated,
  title={Frustrated Ising charge correlations in the kagome metal ScV 6 Sn 6},
  author={Alvarado, SJ Gomez and Pokharel, Ganesh and Ortiz, Brenden R and Paddison, Joseph AM and Sarker, Suchismita and Ruff, Jacob PC and Wilson, Stephen D},
  journal={Phys. Rev. B},
  volume={110},
  number={14},
  pages={L140304},
  year={2024},
  publisher={APS}
}

@article{baral2022real,
  title={{Real-space visualization of short-range antiferromagnetic correlations in a magnetically enhanced thermoelectric}},
  author={Baral, Raju and Christensen, Jacob A and Hamilton, Parker K and Ye, Feng and Chesnel, Karine and Sparks, Taylor D and Ward, Rosa and Yan, Jiaqiang and McGuire, Michael A and Manley, Michael E and others},
  journal={Matter},
  volume={5},
  number={6},
  pages={1853--1864},
  year={2022},
  publisher={Elsevier}
}

@Article{baral2023,
  author    = {Baral, Raju and Abeykoon, AM Milinda and Campbell, Branton J and Frandsen, Benjamin A},
  title     = {{Giant spontaneous magnetostriction in MnTe driven by a novel magnetostructural coupling mechanism}},
  journal   = {Adv. Funct. Mater.},
  year      = {2023},
  volume    = {33},
  number    = {46},
  pages     = {2305247},
  publisher = {Wiley Online Library},
}

@article{boraley2025microscopic,
  title={Microscopic origin of reduced magnetic order in a frustrated metal},
  author={Boraley, X and Stockert, O and Lass, J and Sibille, R and Fjellv{\aa}g, {\O}S and Moody, SH and L{\"a}uchli, AM and Fritsch, Veronika and Mazzone, DG},
  journal={Phys. Rev. Lett.},
  volume={135},
  number={4},
  pages={046702},
  year={2025},
  publisher={APS}
}

@Article{bossini2021femtosecond,
  author    = {Bossini, Davide and Dal Conte, Stefano and Terschanski, Marc and Springholz, Gunther and Bonanni, Alberta and Deltenre, Kira and Anders, F and Uhrig, G{\"o}tz S and Cerullo, Giulio and Cinchetti, Mirko},
  title     = {{Femtosecond phononic coupling to both spins and charges in a room-temperature antiferromagnetic semiconductor}},
  journal   = {Phys. Rev. B},
  year      = {2021},
  volume    = {104},
  number    = {22},
  pages     = {224424},
  publisher = {APS},
}

@article{brout1967molecular,
  title={Molecular field theory, the Onsager reaction field and the spherical model},
  author={Brout, R and Thomas, H},
  journal={Phys. Phys. Fiz.},
  volume={3},
  number={6},
  pages={317},
  year={1967},
  publisher={APS}
}

@article{calder2025investigating,
  title={Investigating Quantum Materials with Half-Polarized Diffraction and magnetic PDF analysis at the HB-2A Neutron Powder Diffractometer},
  author={Calder, Stuart and Baral, Raju and Powell, Matthew and Kolis, Joseph and Zhou, Haidong},
  journal={Struct. Dyn.},
  volume={12},
  number={5\_Supplement},
  pages={A102--A102},
  year={2025},
  publisher={American Crystallographic Association}
}

@article{canals2000mean,
  title={Mean-field study of the disordered ground state in the $\beta$- Mn lattice},
  author={Canals, Benjamin and Lacroix, Claudine},
  journal={Phys. Rev. B},
  volume={61},
  number={17},
  pages={11251},
  year={2000},
  publisher={APS}
}

@article{carlisle2025tuning,
  title={Tuning the magnetic properties of the spin-split antiferromagnet MnTe through pressure},
  author={Carlisle, Edison P and Yumnam, George and Calder, Stuart and Haberl, Bianca and Xiong, Jia-Xin and McGuire, Michael A and Zunger, Alex and Hermann, Rapha{\"e}l P and Frandsen, Benjamin A},
  journal={Phys. Rev. B},
  volume={112},
  number={1},
  pages={014450},
  year={2025},
  publisher={APS}
}

@misc{carlisle2026data,
  title        = {{Data and Analysis Notebooks for Probing Exchange Interactions in Real Space}},
  author       = {Carlisle, Edison P. and Frandsen, Benjamin A.},
  year         = {2026},
  howpublished = {\url{https://zenodo.org/records/22900280}},
  note         = {\textrm{Zenodo}, DOI: 10.5281/zenodo.22900280}
}

@article{cyrot1979new,
  title={New point of view on metastability in spin-glasses},
  author={Cyrot, M},
  journal={Phys. Rev. Lett.},
  volume={43},
  number={2},
  pages={173},
  year={1979},
  publisher={APS}
}

@article{dally2018amplitude,
  title={Amplitude mode in the planar triangular antiferromagnet Na0. 9MnO2},
  author={Dally, Rebecca L and Zhao, Yang and Xu, Zhijun and Chisnell, Robin and Stone, Matthew B and Lynn, Jeffrey W and Balents, Leon and Wilson, Stephen D},
  journal={Nat. Commun.},
  volume={9},
  number={1},
  pages={2188},
  year={2018},
  publisher={Nature Publishing Group UK London}
}

@article{dally2018thermal,
  title={Thermal evolution of quasi-one-dimensional spin correlations within the anisotropic triangular lattice of $\alpha$-NaMnO 2},
  author={Dally, Rebecca L and Chisnell, Robin and Harriger, Leland and Liu, Yaohua and Lynn, Jeffrey W and Wilson, Stephen D},
  journal={Phys. Rev. B},
  volume={98},
  number={14},
  pages={144444},
  year={2018},
  publisher={APS}
}

@article{deng2021high,
  title={{High-performance eco-friendly MnTe thermoelectrics through introducing SnTe nanocrystals and manipulating band structure}},
  author={Deng, Houquan and Lou, Xunuo and Lu, Wenqi and Zhang, Jian and Li, Di and Li, Shuang and Zhang, Qingtang and Zhang, Xuemei and Chen, Xiang and Zhang, Dewei and others},
  journal={Nano Energy},
  volume={81},
  pages={105649},
  year={2021},
  publisher={Elsevier}
}

@article{desai2026machine,
  title={Machine learning magnetic interactions from neutron powder diffraction data},
  author={Desai, Adit S and Cheng, Yongqiang and Paddison, Joseph AM},
  journal={arXiv preprint arXiv:2609.21970},
  year={2026}
}

@article{desgranges2025can,
  title={What can studies of the pair-distribution functions (PDF) contribute to our knowledge about UO2?},
  author={Desgranges, Lionel and Fischer, Henry E and Lander, Gerard H and Baldinozzi, Gianguido},
  journal={MRS Commun.},
  volume={15},
  number={3},
  pages={441--446},
  year={2025},
  publisher={Springer}
}

@Article{dong2018lead,
  author    = {Dong, Jinfeng and Wu, Chao-Feng and Pei, Jun and Sun, Fu-Hua and Pan, Yu and Zhang, Bo-Ping and Tang, Huaichao and Li, Jing-Feng},
  title     = {{Lead-free MnTe mid-temperature thermoelectric materials: facile synthesis, p-type doping and transport properties}},
  journal   = {J. Mater. Chem. C},
  year      = {2018},
  volume    = {6},
  number    = {15},
  pages     = {4265--4272},
  publisher = {Royal Society of Chemistry},
}

@article{dun2021neutron,
  title={Neutron scattering investigation of proposed Kosterlitz-Thouless transitions in the triangular-lattice Ising antiferromagnet TmMgGaO 4},
  author={Dun, Zhiling and Daum, Marcus and Baral, Raju and Fischer, Henry E and Cao, Huibo and Liu, Yaohua and Stone, Matthew B and Rodriguez-Rivera, Jose A and Choi, Eun Sang and Huang, Qing and others},
  journal={Phys. Rev. B},
  volume={103},
  number={6},
  pages={064424},
  year={2021},
  publisher={APS}
}

@article{eastwood1995onsager,
  title={Onsager reaction field theory of a spatially anisotropic Heisenberg model},
  author={Eastwood, Michael P and Logan, David E},
  journal={Phys. Rev. B},
  volume={52},
  number={13},
  pages={9455},
  year={1995},
  publisher={APS}
}

@article{enjalran2004theory,
  title={Theory of paramagnetic scattering in highly frustrated magnets with long-range dipole-dipole interactions: The case of the Tb 2 Ti 2 O 7 pyrochlore antiferromagnet},
  author={Enjalran, Matthew and Gingras, Michel JP},
  journal={Phys. Rev. B, Condens. Matter Mater. Phys.},
  volume={70},
  number={17},
  pages={174426},
  year={2004},
  publisher={APS}
}

@book{fishman2018spin,
  title={Spin-Wave Theory and its Applications to Neutron Scattering and THz Spectroscopy},
  author={Fishman, Randy S and Fernandez-Baca, Jaime A and R{\~o}{\~o}m, Toomas},
  year={2018},
  publisher={Morgan \& Claypool Publishers}
}

@article{frandsen2014magnetic,
  title={Magnetic pair distribution function analysis of local magnetic correlations},
  author={Frandsen, Benjamin and Yang, Xiaohao and Billinge, Simon JL},
  journal={Acta Cryst.},
  volume={70},
  number={1},
  pages={3--11},
  year={2014},
  publisher={International Union of Crystallography}
}

@article{frandsen2015magnetic,
  title={Magnetic structure determination from the magnetic pair distribution function (mPDF): ground state of MnO},
  author={Frandsen, Benjamin A and Billinge, Simon JL},
  journal={Acta Cryst. A},
  volume={71},
  number={3},
  pages={325--334},
  year={2015},
  publisher={International Union of Crystallography}
}

@article{frandsen2020nanoscale,
  title={Nanoscale degeneracy lifting in a geometrically frustrated antiferromagnet},
  author={Frandsen, Benjamin A and Bozin, Emil S and Aza, Eleni and Mart{\'\i}nez, Antonio Fern{\'a}ndez and Feygenson, Mikhail and Page, Katharine and Lappas, Alexandros},
  journal={Phys. Rev. B},
  volume={101},
  number={2},
  pages={024423},
  year={2020},
  publisher={APS}
}

@article{frandsen2022diffpy,
  title={diffpy. mpdf: open-source software for magnetic pair distribution function analysis},
  author={Frandsen, Benjamin A and Hamilton, Parker K and Christensen, Jacob A and Stubben, Eric and Billinge, Simon JL},
  journal={J. Appl. Crystallogr.},
  volume={55},
  number={5},
  pages={1377--1382},
  year={2022},
  publisher={International Union of Crystallography}
}

@article{frandsen2022magnetic,
  title={Magnetic pair distribution function data using polarized neutrons and ad hoc corrections},
  author={Frandsen, Benjamin A and Baral, Raju and Winn, Barry and Garlea, V Ovidiu},
  journal={J. Appl. Phys.},
  volume={132},
  number={22},
  year={2022},
  publisher={AIP Publishing}
}

@article{frandsen2024new,
  title={A new spin on material properties: Local magnetic structure in functional and quantum materials},
  author={Frandsen, Benjamin A and Fischer, Henry E},
  journal={Chem. Mater.},
  volume={36},
  number={19},
  pages={9089--9106},
  year={2024},
  publisher={ACS Publications}
}

@article{furdyna1987magnetic,
  title={Magnetic properties of diluted magnetic semiconductors: a review},
  author={Furdyna, JK and Samarth, N},
  journal={J. Appl. Phys.},
  volume={61},
  number={8},
  pages={3526--3531},
  year={1987},
  publisher={American Institute of Physics}
}

@article{giot2007magnetoelastic,
  title={Magnetoelastic Coupling and Symmetry Breaking in the Frustrated Antiferromagnet $\alpha$-NaMnO 2},
  author={Giot, Maud and Chapon, Laurent C and Androulakis, John and Green, Mark A and Radaelli, Paolo G and Lappas, Alexandros},
  journal={Phys. Rev. Lett.},
  volume={99},
  number={24},
  pages={247211},
  year={2007},
  publisher={APS}
}

@Article{gonzalez2023spontaneous,
  author    = {Gonzalez Betancourt, RD and Zub{\'a}{\v{c}}, Jan and Gonzalez-Hernandez, R and Geishendorf, Kevin and {\v{S}}ob{\'a}{\v{n}}, Zbynek and Springholz, Gunther and Olejn{\'\i}k, Kamil and {\v{S}}mejkal, Libor and Sinova, Jairo and Jungwirth, Tomas and others},
  title     = {{Spontaneous anomalous Hall effect arising from an unconventional compensated magnetic phase in a semiconductor}},
  journal   = {Phys. Rev. Lett.},
  year      = {2023},
  volume    = {130},
  number    = {3},
  pages     = {036702},
  publisher = {APS},
}

@book{gradshteyn2015,
    author = {I.S. Gradshteyn and I.M. Ryzhik},
    title = {Table of Integrals, Series, and Products},
    publisher = {Academic Press},
    edition = {8th},
    year = {2015},
    address = {Boston},
    isbn = {978-0-12-384933-5},
    doi = {https://doi.org/10.1016/B978-0-12-384933-5.00003-5},
    url = {https://www.sciencedirect.com/science/article/pii/B9780123849335000035}
}

@article{hatt2025cluster,
  title={Cluster spin glass correlations and dynamics in Zn$_{0.5}$Mn$_{0.5}$Te},
  author={Hatt, Sabrina R and Shaw, Camille and Zappala, Emma and Baral, Raju and Calder, Stuart and Morris, Gerald D and Ortiz, Brenden R and Chesnel, Karine and Frandsen, Benjamin A},
  journal={Phys. Rev. B},
  volume={112},
  number={14},
  pages={144440},
  year={2025},
  publisher={APS}
}

@article{hohlwein2002magnetic,
  title={Magnetic interactions and Onsager reaction field from paramagnetic diffuse neutron scattering},
  author={Hohlwein, D},
  journal={Appl. Phys. A},
  volume={74},
  number={Suppl 1},
  pages={s740--s742},
  year={2002},
  publisher={Springer}
}

@article{johnston1961mnte,
  title={The MnTe-GeTe phase diagram},
  author={Johnston, WD and Sestrich, DE},
  journal={J. Inorg. Nucl. Chem.},
  volume={19},
  number={3-4},
  pages={229--236},
  year={1961},
  publisher={Elsevier}
}

@article{kanada2002neutron,
  title={Neutron scattering study of the spin correlation in the spin ice system Ho2Ti2O7},
  author={Kanada, Masaki and Yasui, Yukio and Kondo, Yasuyuki and Iikubo, Satoshi and Ito, Masafumi and Harashina, Hiroshi and Sato, Masatoshi and Okumura, Hajime and Kakurai, Kazuhisa and Kadowaki, Hiroaki},
  journal={J. Phys. Soc. Jpn.},
  volume={71},
  number={1},
  pages={313--318},
  year={2002},
  publisher={The Physical Society of Japan}
}

@article{kodama2017alternative,
  title={Alternative equation on magnetic pair distribution function for quantitative analysis},
  author={Kodama, Katsuaki and Ikeda, Kazutaka and Shamoto, Shin-ichi and Otomo, Toshiya},
  journal={J. Phys. Soc. Jpn.},
  volume={86},
  number={12},
  pages={124708},
  year={2017},
  publisher={The Physical Society of Japan}
}

@article{krempasky2024altermagnetic,
  title={Altermagnetic lifting of Kramers spin degeneracy},
  author={Krempask{\`y}, Juraj and {\v{S}}mejkal, L and D’souza, SW and Hajlaoui, M and Springholz, G and Uhl{\'\i}{\v{r}}ov{\'a}, K and Alarab, F and Constantinou, PC and Strocov, V and Usanov, D and others},
  journal={Nat.},
  volume={626},
  number={7999},
  pages={517--522},
  year={2024},
  publisher={Nature Publishing Group UK London}
}

@Article{kriegner2016multiple,
  author    = {Kriegner, Dominik and V{\`y}born{\`y}, K and Olejn{\'\i}k, K and Reichlov{\'a}, H and Nov{\'a}k, V and Marti, X and Gazquez, J and Saidl, V and N{\v{e}}mec, P and Volobuev, VV and others},
  title     = {{Multiple-stable anisotropic magnetoresistance memory in antiferromagnetic MnTe}},
  journal   = {Nat. Commun.},
  year      = {2016},
  volume    = {7},
  number    = {1},
  pages     = {11623},
  publisher = {Nature Publishing Group UK London},
}

@article{kubo1952spin,
  title={The spin-wave theory of antiferromagnetics},
  author={Kubo, Ryogo},
  journal={Phys. Rev.},
  volume={87},
  number={4},
  pages={568},
  year={1952},
  publisher={APS}
}

@article{kurniawan2026information,
  title={An information-matching approach to optimal experimental design and active learning},
  author={Kurniawan, Yonatan and Neilsen, Tracianne B and Francis, Benjamin L and Stankovic, Alex M and Wen, Mingjian and Nikiforov, Ilia and Tadmor, Ellad B and Bulatov, Vasily V and Lordi, Vincenzo and Transtrum, Mark K},
  journal={Appl. Phys. Lett.},
  volume={128},
  number={6},
  year={2026},
  publisher={AIP Publishing}
}

@Article{lee2024broken,
  author    = {Lee, Suyoung and Lee, Sangjae and Jung, Saegyeol and Jung, Jiwon and Kim, Donghan and Lee, Yeonjae and Seok, Byeongjun and Kim, Jaeyoung and Park, Byeong Gyu and {\v{S}}mejkal, Libor and others},
  title     = {{Broken Kramers degeneracy in altermagnetic MnTe}},
  journal   = {Phys. Rev. Lett.},
  year      = {2024},
  volume    = {132},
  number    = {3},
  pages     = {036702},
  publisher = {APS},
}

@article{logan1995onsager,
  title={Onsager reaction field theory of the Heisenberg model},
  author={Logan, DE and Szczech, YH and Tusch, MA},
  journal={Europhys. Lett.},
  volume={30},
  number={5},
  pages={307--312},
  year={1995}
}

@Article{mazin2023altermagnetism,
  author    = {Mazin, II},
  title     = {{Altermagnetism in MnTe: Origin, predicted manifestations, and routes to detwinning}},
  journal   = {Phys. Rev. B},
  year      = {2023},
  volume    = {107},
  number    = {10},
  pages     = {L100418},
  publisher = {APS},
}

@article{momma2011vesta,
  title={VESTA 3 for three-dimensional visualization of crystal, volumetric and morphology data},
  author={Momma, Koichi and Izumi, Fujio},
  journal={J. Appl. Crystallogr.},
  volume={44},
  number={6},
  pages={1272--1276},
  year={2011},
  publisher={International Union of Crystallography}
}

@Article{mu2019phonons,
  author    = {Mu, Sai and Hermann, Rapha{\"e}l P and Gorsse, St{\'e}phane and Zhao, Huaizhou and Manley, Michael E and Fishman, Randy S and Lindsay, Lucas},
  title     = {{Phonons, magnons, and lattice thermal transport in antiferromagnetic semiconductor MnTe}},
  journal   = {Phys. Rev. Mater.},
  year      = {2019},
  volume    = {3},
  number    = {2},
  pages     = {025403},
  publisher = {APS},
}

@article{murta2022one,
  title={One-to-one correspondence between thermal structure factors and coupling constants of general bilinear Hamiltonians},
  author={Murta, Bruno and Fern{\'a}ndez-Rossier, J},
  journal={Physical Review E},
  volume={105},
  number={6},
  pages={L062101},
  year={2022},
  publisher={APS}
}

@article{nozieres1982onsager,
  title={Onsager anisotropies in Heisenberg spin glasses},
  author={Nozieres, P},
  journal={J. Phys., Lett.},
  volume={43},
  number={15},
  pages={543--549},
  year={1982},
  publisher={Les Editions de Physique}
}

@article{nuttall2023quantitative,
  title={Quantitative investigation of the short-range magnetic correlations in the candidate quantum spin liquid NaYbO 2},
  author={Nuttall, Kristina Michelle and Suggs, Christiana Z and Fischer, Henry E and Bordelon, Mitchell M and Wilson, Stephen D and Frandsen, Benjamin A},
  journal={Phys. Rev. B},
  volume={108},
  number={14},
  pages={L140411},
  year={2023},
  publisher={APS}
}

@article{paddison2020scattering,
  title={Scattering signatures of bond-dependent magnetic interactions},
  author={Paddison, Joseph AM},
  journal={Phys. Rev. Lett.},
  volume={125},
  number={24},
  pages={247202},
  year={2020},
  publisher={APS}
}

@article{paddison2023spinteract,
  title={Spinteract: a program to refine magnetic interactions to diffuse scattering data},
  author={Paddison, Joseph AM},
  journal={J. Phys.: Condens. Matter},
  volume={35},
  number={49},
  pages={495802},
  year={2023},
  publisher={IOP Publishing}
}

@article{paddison2026kitaev,
  title={Kitaev interactions of the spin--orbit coupled magnet UO2},
  author={Paddison, Joseph AM and Desgranges, Lionel and Baldinozzi, Gianguido and Lander, Gerard H and Fischer, Henry E},
  journal={J. Phys. Condens. Matter},
  volume={38},
  number={16},
  pages={165801},
  year={2026},
  publisher={IOP Publishing}
}

@article{pitcairn2024controlling,
  title={Controlling noncollinear ferromagnetism in van der Waals metal--organic magnets},
  author={Pitcairn, Jem and Ongkiko, Mario Antonio and Iliceto, Andrea and Speakman, Peter J and Calder, Stuart and Cochran, Malcolm J and Paddison, Joseph AM and Liu, Cheng and Argent, Stephen P and Morris, Andrew J and others},
  journal={J. Am. Chem. Soc.},
  volume={146},
  number={28},
  pages={19146--19159},
  year={2024},
  publisher={ACS Publications}
}

@Article{polash2021understanding,
  author    = {Polash, Md Mobarak Hossain and Moseley, Duncan and Zhang, Junjie and Hermann, Raphael P and Vashaee, Daryoosh},
  title     = {{Understanding and design of spin-driven thermoelectrics}},
  journal   = {Cell Rep. Phys. Sci.},
  year      = {2021},
  volume    = {2},
  number    = {11},
  publisher = {Elsevier},
}

@article{potashnikov2025magnetic,
  title={Magnetic interactions in (Mo 2/3 R 1/3) 2 AlC rare-earth i-MAX phases (R= Nd, Sm, Gd, Tb, Dy, Ho, and Er) from first principles and experiment},
  author={Potashnikov, Daniel and Caspi, El'ad Nisan and Tao, Quanzheng and Rosen, Johanna and Cheptyakov, Denis and Ritter, Clemens and Di{\'e}guez, Oswaldo},
  journal={Phys. Rev. B},
  volume={112},
  number={6},
  pages={064407},
  year={2025},
  publisher={APS}
}

@article{quintanilla2022relationship,
  title={Relationship between the wave function of a magnet and its static structure factor},
  author={Quintanilla, Jorge},
  journal={Physical Review B},
  volume={106},
  number={10},
  pages={104435},
  year={2022},
  publisher={APS}
}

@article{ramirez2025short,
  title={Short-range order and hidden energy scale in geometrically frustrated magnets},
  author={Ramirez, Arthur P and Syzranov, SV},
  journal={Mater. Adv.},
  volume={6},
  number={4},
  pages={1213--1229},
  year={2025},
  publisher={Royal Society of Chemistry}
}

@article{reimers1991mean,
  title={Mean-field approach to magnetic ordering in highly frustrated pyrochlores},
  author={Reimers, JN and Berlinsky, AJ and Shi, A-C},
  journal={Phys. Rev. B},
  volume={43},
  number={1},
  pages={865},
  year={1991},
  publisher={APS}
}

@article{reimers1992diffuse,
  title={Diffuse-magnetic-scattering calculations for frustrated antiferromagnets},
  author={Reimers, Jan N},
  journal={Phys. Rev. B},
  volume={46},
  number={1},
  pages={193},
  year={1992},
  publisher={APS}
}

@Article{ren2017synergistic,
  author    = {Ren, Yangyang and Yang, Junyou and Jiang, Qinghui and Zhang, Dan and Zhou, Zhiwei and Li, Xin and Xin, Jiwu and He, Xu},
  title     = {{Synergistic effect by Na doping and S substitution for high thermoelectric performance of p-type MnTe}},
  journal   = {J. Mater. Chem. C},
  year      = {2017},
  volume    = {5},
  number    = {21},
  pages     = {5076--5082},
  publisher = {Royal Society of Chemistry},
}

@article{roth2018model,
  title={Model-free reconstruction of magnetic correlations in frustrated magnets},
  author={Roth, Nikolaj and May, Andrew F and Ye, Feng and Chakoumakos, Bryan C and Iversen, Bo Brummerstedt},
  journal={IUCrJ},
  volume={5},
  number={4},
  pages={410--416},
  year={2018},
  publisher={International Union of Crystallography}
}

@Article{vsmejkal2022beyond,
  author    = {{\v{S}}mejkal, Libor and Sinova, Jairo and Jungwirth, Tomas},
  title     = {{Beyond conventional ferromagnetism and antiferromagnetism: A phase with nonrelativistic spin and crystal rotation symmetry}},
  journal   = {Phys. Rev. X},
  year      = {2022},
  volume    = {12},
  number    = {3},
  pages     = {031042},
  publisher = {APS},
}

@book{squires1996introduction,
  title={Introduction to the theory of thermal neutron scattering},
  author={Squires, Gordon Leslie},
  year={1996},
  publisher={Dover Publications}
}

@article{stock2009one,
  title={One-dimensional magnetic fluctuations in the spin-2 triangular lattice $\alpha$-NaMnO 2},
  author={Stock, Christopher and Chapon, Laurent C and Adamopoulos, O and Lappas, A and Giot, M and Taylor, JW and Green, MA and Brown, <? format?> CM and Radaelli, PG},
  journal={Phys. Rev. Lett.},
  volume={103},
  number={7},
  pages={077202},
  year={2009},
  publisher={APS}
}

@article{szusz_prb06,
  title = {{Spin-wave measurements on hexagonal MnTe NiAs-type structure by inelastic neutron scattering}},
  author = {Szuszkiewicz, W. and Dynowska, E. and Witkowska, B. and Hennion, B.},
  journal = {Phys. Rev. B},
  volume = {73},
  issue = {10},
  pages = {104403},
  numpages = {7},
  year = {2006},
  month = {Mar},
  publisher = {American Physical Society},
  doi = {10.1103/PhysRevB.73.104403},
  url = {https://link.aps.org/doi/10.1103/PhysRevB.73.104403}
}

@Article{walther1967,
  author  = {K. Walther},
  title   = {{Ultrasonic relaxation at the Néel temperature and nuclear acoustic resonance in MnTe}},
  journal = {Solid State Commun.},
  year    = {1967},
  volume  = {5},
  number  = {5},
  pages   = {399-403},
  issn    = {0038-1098},
  doi     = {https://doi.org/10.1016/0038-1098(67)90784-3},
  url     = {https://www.sciencedirect.com/science/article/pii/0038109867907843},
}

@article{wysin2000onsager,
  title={Onsager reaction-field theory for magnetic models on diamond and hcp lattices},
  author={Wysin, GM},
  journal={Phys. Rev. B},
  volume={62},
  number={5},
  pages={3251},
  year={2000},
  publisher={APS}
}

@article{xiong2023lattice,
  title={{Lattice distortions and multiple valence band convergence contributing to high thermoelectric performance in MnTe}},
  author={Xiong, Wenjie and Wang, Zhichao and Zhang, Xuemei and Wang, Chong and Yin, Liangcao and Gong, Yaru and Zhang, Qingtang and Li, Shuang and Liu, Qingfeng and Wang, Peng and others},
  journal={Small},
  volume={19},
  number={6},
  pages={2206058},
  year={2023},
  publisher={Wiley Online Library}
}

@Article{xu2017performance,
  author    = {Xu, Yidong and Li, Wen and Wang, Chen and Li, Juan and Chen, Zhiwei and Lin, Siqi and Chen, Yue and Pei, Yanzhong},
  title     = {{Performance optimization and single parabolic band behavior of thermoelectric MnTe}},
  journal   = {J. Mater. Chem. A},
  year      = {2017},
  volume    = {5},
  number    = {36},
  pages     = {19143--19150},
  publisher = {Royal Society of Chemistry},
}

@article{zhang2021classical,
  title={Classical spin dynamics based on SU (N) coherent states},
  author={Zhang, Hao and Batista, Cristian D},
  journal={Phys. Rev. B},
  volume={104},
  number={10},
  pages={104409},
  year={2021},
  publisher={APS}
}

@Article{zheng2019paramagnon,
  author    = {Zheng, Yuanhua and Lu, Tianqi and Polash, Md MH and Rasoulianboroujeni, Morteza and Liu, Ning and Manley, Michael E and Deng, Yuan and Sun, PJ and Chen, XL and Hermann, Raphael P and others},
  title     = {{Paramagnon drag in high thermoelectric figure of merit Li-doped MnTe}},
  journal   = {Sci. Adv.},
  year      = {2019},
  volume    = {5},
  number    = {9},
  pages     = {eaat9461},
  publisher = {American Association for the Advancement of Science},
}

@article{zorko2008magnetic,
  title={Magnetic interactions in $\alpha$-Na Mn O 2: Quantum spin-2 system on a spatially anisotropic two-dimensional triangular lattice},
  author={Zorko, Andrej and El Shawish, Samir and Ar{\v{c}}on, Denis and Jagli{\v{c}}i{\'c}, Zvonko and Lappas, Alexandros and van Tol, Hans and Brunel, Louis Claude},
  journal={Phys. Rev. B, Condens. Matter Mater. Phys.},
  volume={77},
  number={2},
  pages={024412},
  year={2008},
  publisher={APS}
}

@Article{zulkifal2023multiple,
  author    = {Zulkifal, Shahzada and Wang, Zhichao and Zhang, Xuemei and Siddique, Suniya and Yu, Yuan and Wang, Chong and Gong, Yaru and Li, Shuang and Li, Di and Zhang, Yongsheng and others},
  title     = {{Multiple valence bands convergence and localized lattice engineering lead to superhigh thermoelectric figure of merit in MnTe}},
  journal   = {Adv. Sci.},
  year      = {2023},
  volume    = {10},
  number    = {17},
  pages     = {2206342},
  publisher = {Wiley Online Library},
}

\appendix
\section{Fourier Transform of $\langle S_\mu(\mathbf{q})S_\mu(\mathbf{-q}) \rangle$}
\label{sec:appendixFT}

Here, we provide a derivation of Equations~\ref{eq:SpinCorrMuAniso} and~\ref{eq:DampingMatrix}. We start with the expression for $\langle S_\mu(\mathbf{q})S_\mu(\mathbf{-q}) \rangle$ from Equation~\ref{eq:SpinCorrQ} and perform a Taylor expansion of $J_\mu(\mathbf{q)}$ around a peak position $\mathbf{q}_0$ in all three components of $\mathbf{q}$.
\begin{multline}
    \langle S_\mu(\mathbf{q})S_\mu(\mathbf{-q}) \rangle = T\frac{\chi_0}{1-\chi_0(J_\mu(\mathbf{q})-\lambda)} \\
    \approx T\frac{\chi_0}{1-\chi_0(J_\mu(\mathbf{q}_0) + \frac{1}{2}\sum_{i,j} C_{ij}\delta q_i \delta q_j-\lambda)},
\end{multline}
where $\delta q_i = q_i-q_{0,i}$. From here, we can arrive at $\langle S_\mu(0)S_\mu(\mathbf{r}) \rangle$ through a 3D Fourier transform:
\begin{multline}
    \langle S_\mu(0)S_\mu(\mathbf{r}) \rangle = \frac{V_\mathrm{UC}}{(2\pi)^{3}}\int \mathrm{d}\mathbf{q}^3 T\frac{\chi_0e^{-i \mathbf{q}\cdot\mathbf{r}}}{1-\chi_0(J_\mu(\mathbf{q})-\lambda)} \\
    \approx \frac{V_\mathrm{UC}}{(2\pi)^{3}}\int \mathrm{d}\mathbf{q}^3 T\frac{e^{-i \mathbf{q}\cdot\mathbf{r}}}{C_0 - \frac{1}{2}\sum_{i,j} C_{ij}\delta q_i \delta q_j} ,
\end{multline}
where $C_{ij} = \partial_{q_i}\partial_{q_j} J_\mu(\mathbf{q}_0)$, $C_0 = 1/\chi_0 + \lambda - J_\mu(\mathbf{q}_0)$, and the integration is across the first Brillouin zone, the volume of which is equal to $(2\pi)^3/V_\mathrm{UC}$. Now, we can integrate across our first reciprocal space dimension. The integration becomes easier if we extend the bounds of integration from $-\infty$ to $\infty$. This approximation is valid because we expect $\langle S_\mu(\mathbf{q})S_\mu(\mathbf{-q}) \rangle$ to be significantly small as close as one reciprocal lattice vector away from $\mathbf{q}_0$. Then, this integral is equivalent to performing the Fourier transform of a Lorentzian.
\begin{widetext}
\begin{multline}
    \langle S_\mu(0)S_\mu(\mathbf{r}) \rangle \approx \frac{V_\mathrm{UC}}{(2\pi)^{3}} T \int \mathrm{d}q_2\mathrm{d}q_3 e^{-i(q_2 y + q_3 z)} \int \mathrm{d}q_1 \frac{1}{C_0 - \frac{1}{2}\sum_{i,j} C_{ij}\delta q_i\delta q_j} e^{-i q_1 x}, \\
    = \frac{\sqrt{2}\pi V_\mathrm{UC}Te^{-iq_{0,1}x}}{(2\pi)^{3}\sqrt{-C_{11}}} \int\mathrm{d}q_2\mathrm{d}q_3 e^{-i (q_2y + q_3 z -\frac{C_{12}\delta q_2 + C_{13}\delta q_3}{C_{11}}x)} \frac{\exp\left(-\sqrt{\frac{C_0 - \frac{1}{2}\sum_{i>1,j>1} C_{ij}\delta q_i\delta q_j + \frac{(C_{12}\delta q_2+C_{13}\delta q_3)^2}{2C_{11}}}{-C_{11}/2}x^2}\right)}{\sqrt{C_0 - \frac{1}{2}\sum_{i>1,j>1} C_{ij}\delta q_i\delta q_j + \frac{(C_{12}\delta q_2+C_{13}\delta q_3)^2}{2C_{11}}}}.
\end{multline}
\end{widetext}

For a 1D magnet, $J_\mu(\mathbf{q)}$ will only be dependent on one reciprocal space dimension. Here, that is equivalent to a non-zero $C_{11}$ and all other elements of $C_{ij}=0$. We set $y$ and $z$ to zero, drop the integration over $q_2$ and $q_3$, and replace the volume of the Brillouin zone, $(2\pi)^3/V_\mathrm{UC}$, with a length, $2\pi/L_\mathrm{UC}$, where $L_\mathrm{UC}$ is the length of the unit cell. Then, for a 1D magnet
\begin{multline}
    \langle S_\mu(0)S_\mu(x) \rangle_\mathrm{1D} \\
    \approx \frac{L_\mathrm{UC}T e^{-iq_{0}x}}{\sqrt{-2\partial_q^2J_\mu(q_0)}} 
    \frac{\exp(-x\sqrt{2\frac{1/\chi_0 + \lambda - J_\mu(q_0)}{-\partial_q^2J_\mu(q_0)}})}{\sqrt{1/\chi_0 + \lambda - J_\mu(q_0)}}.
    \label{eq:SpinCorr1D}
\end{multline}

For the next integral, we proceed by using formula 3.914.4 from Ref.~\cite{gradshteyn2015}:
\begin{equation}
    \int_0^\infty \frac{e^{-\beta \sqrt{\gamma^2 + \xi^2}}}{\sqrt{\gamma^2 + \xi^2}}\cos(b\xi) \mathrm{d}\xi = K_0(\gamma \sqrt{\beta^2+b^2}),
\end{equation}
where $K_0(z)$ is the modified Bessel function of the second kind. The equation
\begin{equation}
    \int_{-\infty}^\infty \frac{e^{-\beta \sqrt{\gamma^2 + x^2}}}{\sqrt{\gamma^2 + x^2}}e^{-ibx} \mathrm{d}x = 2K_0(\gamma \sqrt{\beta^2+b^2}),
    \label{eq:integral1}
\end{equation}
follows from the previous formula.
We can make the substitutions
\begin{align}
    \gamma^2 &\rightarrow C_0 + \left [ \frac{1}{2}(\frac{C_{13}^2}{C_{11}} - C_{33}) -\frac{(C_{12}C_{13}-C_{11}C_{23})^2}{2C_{11}(C_{12}^2-C_{11}C_{22})}\right ] \delta q_3^2,\\
    \xi^2 &\rightarrow \sqrt{\frac{1}{2}(\frac{C_{12}^2}{C_{11}} - C_{22})} \left(\delta q_2 + \frac{C_{12}C_{13}-C_{11}C_{23}}{C_{12}^2-C_{11}C_{22}}\delta q_3 \right ), \\
    \beta^2 &\rightarrow -2x^2/C_{11},\\
    b &\rightarrow (y - \frac{C_{12}}{C_{11}}x) \sqrt{\frac{2 C_{11}}{C_{12}^2-C_{11}C_{22}}}.
\end{align}
Again, we extend our bounds of integration in our spin correlation function from $-\infty$ to $\infty$ to match Equation~\ref{eq:integral1}. Thus, our spin correlation function becomes
\begin{widetext}
\begin{multline}
    \langle S_\mu(0)S_\mu(\mathbf{r}) \rangle \approx \frac{2 V_\mathrm{UC}T}{(2\pi)^2}\frac{e^{-iq_{0,1}x}e^{-i q_{0,2}y}}{\sqrt{C_{11}C_{22}- C_{12}^2}}  \int\mathrm{d}q_3 e^{-i q_3 z} e^{i \frac{(C_{12}C_{13}-C_{11}C_{23})y + (C_{12}C_{23}-C_{22}C_{13})x}{C_{12}^2-C_{11}C_{22}} \delta q_3} \\
    \times K_0 \left ( \sqrt{C_0 + \left [ \frac{1}{2}(\frac{C_{13}^2}{C_{11}} - C_{33}) -\frac{(C_{12}C_{13}-C_{11}C_{23})^2}{2C_{11}(C_{12}^2-C_{11}C_{22})}\right ] \delta q_3^2} \sqrt{\frac{2C_{11}y^2-4C_{12}xy+2C_{22}x^2}{C_{12}^2-C_{11}C_{22}}} ~\right ).
\end{multline}
\end{widetext}
For a 2D magnet, we take the same steps we did before for the 1D magnet, but now we only exclude $q_3$. Our spin correlation function for a 2D magnet is
\begin{equation}
    \langle S_\mu(0)S_\mu(\mathbf{r}) \rangle_\mathrm{2D} \approx \frac{A_\mathrm{UC}Te^{-\mathbf{q_0}\cdot\mathbf{r}}}{\pi\sqrt{\det C}}
    K_0 \left (\sqrt{\mathbf{r}^TD\mathbf{r}} ~\right ),
    \label{eq:SpinCorr2D}
\end{equation}
where $A_\mathrm{UC}$ is the area of the unit cell,$D = -2 C_0 C^{-1}$, and $C$ is now only a $2\times2$ matrix (we exclude $C_{13}$, $C_{23}$, and $C_{33}$).

For the final integration, we use formula 6.677.5 from Ref.~\cite{gradshteyn2015}:
\begin{multline}
    \int_0^\infty K_0(a \sqrt{\xi^2 + \beta^2})\cos(\gamma \xi)\mathrm{d}x \\
    = \frac{\pi}{2\sqrt{a^2+\gamma^2}}\exp(-\beta \sqrt{a^2 + \gamma^2}),
\end{multline}
which can be expanded to
\begin{multline}
    \int_{-\infty}^\infty K_0(a \sqrt{\xi^2 + \beta^2})e^{i\gamma \xi}\mathrm{d}\xi \\
    = \frac{\pi}{\sqrt{a^2+\gamma^2}}\exp(-\beta \sqrt{a^2 + \gamma^2}).
\end{multline}

We make the substitutions
\begin{align}
    a &\rightarrow \sqrt{\frac{2C_{11}y^2-4C_{12}xy+2C_{22}x^2}{C_{12}^2-C_{11}C_{22}}},\\
    \beta &\rightarrow \sqrt{C_0},\\
    \gamma &\rightarrow \frac{\left [(y - \frac{C_{12}}{C_{11}}x) (\frac{C_{12}C_{13}-C_{11}C_{23}}{C_{12}^2-C_{11}C_{22}}) + \frac{C_{13}}{C_{11}} x - z\right ]}{\sqrt{\frac{1}{2}(\frac{C_{13}^2}{C_{11}} - C_{33}) -\frac{(C_{12}C_{13}-C_{11}C_{23})^2}{2C_{11}(C_{12}^2-C_{11}C_{22})}}},\\
    \xi &\rightarrow \delta q_3 \sqrt{\frac{1}{2}(\frac{C_{13}^2}{C_{11}} - C_{33}) -\frac{(C_{12}C_{13}-C_{11}C_{23})^2}{2C_{11}(C_{12}^2-C_{11}C_{22})}},
\end{align}
and again extend our bounds of integration from $-\infty$ to $\infty$ to arrive at our spin correlation function in three dimensions:
\begin{equation}
    \langle S_\mu(0)S_\mu(\mathbf{r}) \rangle \approx \frac{V_\mathrm{UC}T}{2\pi}\frac{e^{-i\mathbf{q}_0\cdot\mathbf{r}}}{\sqrt{-\det C}} 
    \frac{\exp(- \sqrt{\mathbf{r}^T D~\mathbf{r}})}
    {\sqrt{-\mathbf{r}^T C^{-1}~\mathbf{r}}},
\end{equation}
where, again, $D = -2 C_0 C^{-1}$, but now with all components of $C$ included.

\vspace{12cm}

\end{document}